\documentclass[fleqn,usenatbib]{mnras}

\usepackage{newtxtext,newtxmath}

\usepackage[T1]{fontenc}

\DeclareRobustCommand{\VAN}[3]{#2}
\let\VANthebibliography\thebibliography
\def\thebibliography{\DeclareRobustCommand{\VAN}[3]{##3}\VANthebibliography}

\usepackage{graphicx}	
\usepackage{amsmath}	

\title[CIRs, impacts with exoplanets]{Corotating Interaction Regions (CIRs): impacts with exoplanets}

\author[Rose F. P. Waugh et al.]{
Rose F.P. Waugh,$^{1}$\thanks{E-mail: rw47@st-andrews.ac.uk (RFPW)}
Moira M. Jardine,$^{1}$
\\
$^{1}$School of Physics and Astronomy, University of St Andrews, North Haugh, St Andrews, Fife, Scotland, KY16 9SS
}

\date{Accepted XXX. Received YYY; in original form ZZZ}

\pubyear{\the\year{}}

\begin{document}
\label{firstpage}
\pagerange{\pageref{firstpage}--\pageref{lastpage}}
\maketitle

\begin{abstract}
Corotating interaction regions (CIRs) are compressions that form in stellar winds when streams of different speeds collide. They form an Archimedean spiral around the star and can compress any exoplanetary magnetospheres they impact. They may also steepen into shocks capable of accelerating particles to high energies. We model the frequency and strength of these CIRS for stars of spectral types F-M. We show that the minimum radius, $r_{CIR}=\Delta \phi u_{slow}/\Omega$, at which CIRs form varies strongly with the rotation rate (and hence age) of the star. For some exoplanets, such as those in Earth or Mars orbits, CIRs can form within the exoplanet's orbit at all stellar rotation rates, depending on the angular size of the fast wind segment ($\Delta \phi$). These exoplanets will experience CIR impacts at all stellar ages. However, for closer-in orbits such as Mercury or Venus, this may only be the case at higher stellar rotation rates. Both the frequency and impact of CIRs depend on the stellar rotation rate. For exoplanets with $P_{orbit}\gg P_*$, CIR impacts lasting for a time $\Delta t$ raise the exoplanetary outflow rate by a factor $R$. If $P_*\leq N\Delta t$ the CIR pulses overlap in time, whereas if $N\Delta t < P_* \leq N\Delta t(R+1)$, the planet experiences discrete pulses of compression and relaxation and the CIR-related outflow is more than 50$\%$ of the total. For $P_* > N\Delta t(R+1)$ the pulses are less frequent, and contribute less than $50\%$ of the total outflow.

\end{abstract}

\begin{keywords}
stars: activity -- stars: low-mass -- stars: solar-type -- stars: winds, outflows -- planet-star interactions
\end{keywords}



\section{Introduction}

Over much of their lives, stars like the Sun (spectral types F-M) lose both mass and angular momentum in a hot, magnetically-channelled wind~\citep{Parker1958,WeberDavis67}.  This continuous stream of particles, primarily ionised hydrogen and electrons, flows out from the stellar surface along the open magnetic field lines. The winds of stars have important consequences for both the stars themselves and any close-in planets that may be orbiting them~\citep{Zendejas2010,See2014,Vidotto2015,2024ApJ...976...65R,2020IJAsB..19..136A}. 
The angular momentum loss from the star reduces the stellar spin as the star ages,  altering the magnetic field regeneration via the stellar dynamo. This influences other stellar properties such as the elemental abundances and the coronal temperature~\citep{VidottoGregory2014,Reiners2022,Kraft1967,Skumanich1972,Reames2024}.
 The winds of stars are also an important part of the planetary environment, being responsible for geomagnetic storms on Earth (~\cite{Tsurutani1997,Watari2023,Liu2019} and references within). They may also be able to strip atmospheres from exoplanets~\citep{Harbach2021,Holmstrom2008,Khodachenko2012,Kislyakova2014,Alvarado2016}. The shape and compression of planetary magnetospheres are determined by the planetary magnetic pressure and the relative dynamic pressure of the wind to the escaping atmosphere. The interaction between these two has been well studied~\citep{Schneiter2007,Caro2014,DaleyYates2017,McCann2019,Vidotto2020}. Additionally, the effect of these winds may be strengthened by eruptive flares and coronal mass ejections (CMEs). Combined with high XUV radiation, these additional processes shape the space weather that could cause exoplanetary atmosphere loss and chemical changes~\citep{Linsky2019,Airapetian2020}. These chemical changes in an exoplanet's atmosphere occur when high-energy particles interact with the upper atmosphere of an exoplanet, altering the composition through mechanisms such as free-radical generation.\\
\begin{figure}
    \centering \includegraphics[width=0.8\columnwidth]{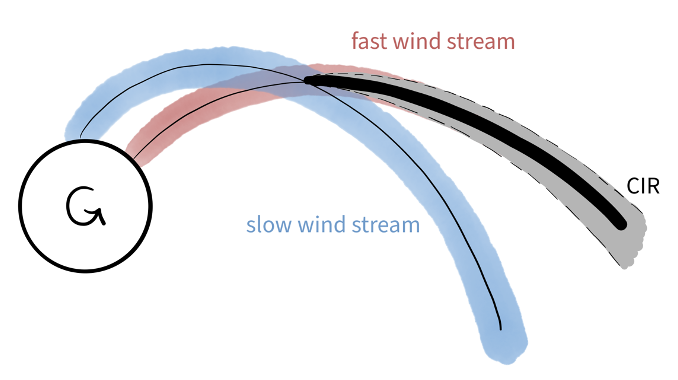}
    \caption{A cartoon showing the collision of a fast (red) and slow (blue) wind stream, leading to the formation of a CIR. The system is viewed from a point above the rotation pole.}
\label{fig:cartoon1}
\end{figure}

The solar wind is composed of two parts: the fast wind, originating from open magnetic field regions, and the slow wind that escapes intermittently and originates from reconnecting closed field regions such as above helmet streamers. The magnetic structure of other low mass stars is not identical to that of the Sun. Surface magnetograms of other stars can be obtained through Zeeman Doppler Imaging~\citep{DonatiLandstreet2009} which can be used to model the coronal magnetic structure and wind evolution~\citep{Vidotto2009,Matt2012,Reville2015,See2015,2023MNRAS.524.2042E}. Since the structure of the solar wind is determined by the solar magnetic field, the different field structures of other stars (~\cite{Jeffers2023} and references within), should lead to variations in their fast and slow wind distributions.\\

A stream interaction region (SIR) is formed by the interaction of a fast solar wind stream with a slow wind stream. This interaction generates a compressed region of plasma along the leading edge of the fast wind stream. These interaction regions often last for multiple solar rotations, in which case they may be referred to as corotating interaction regions (CIRs), while in other cases CIR and SIR are used interchangeably. As seen in Figure~\ref{fig:cartoon1}, the wind streams trace an Archimedes spiral, with the slow wind streams tracing a tighter spiral than the fast wind. Due to the frozen-in-flux theory \citep{alfven1943}, these streams cannot mix. Rather, upon interaction, the fast wind deflects the slower material upstream whilst the slow material deflects the fast material downstream. This compression forms the CIR, as the plasma density and magnetic field intensity rises.
Thus, the CIR forms along the leading edge of the fast stream, where the fast moving material has caught up and collided with the slower wind. In cases where the fast wind is supersonic upon reaching the slow wind, a \textit{forward shock} is formed ahead of the CIR and a \textit{reverse shock} is formed behind. 
The steepening of CIRs into shocks has consequences for orbiting planets, since these shocks are responsible for generating high energy particles that rain back towards the inner Solar System~\citep{RichardsonReview}. However,~\cite{Giacalone2002} showed that a fully formed shock is not necessary for particle acceleration. CIRs can be found at all phases of the solar cycle, although they are most noteworthy during the declining and minimum phases of the solar activity cycle (\cite{Richardson2004} and references within). While this form of space weather is less energetic than CMEs or flares, the persistence of CIRs across the solar cycle (and indeed increase at cycle minimum) results in a larger cumulative effect for Earth through the energetic electron fluxes that CIR shocks may produce \citep{Asikainen2016}. \\

CIRs are important for exoplanets not only because of the highly energetic particles they generate, but also because of their ability to compress the exoplanetary magnetosphere~\citep{Smith1981,Borovsky2006}. Compression of the magnetosphere of planets within our Solar System from solar CIRs has been well studied~\citep{Edberg2011,Dunn2020}. The compression of magnetospheres of exoplanets by stellar CIRs has been modelled by~\cite{Airapetian2021}, who found that the CIRs in the wind of the young solar analogue $\kappa$ Ceti could raise the dynamic pressure of the stellar wind at 1AU by a factor of 1300, enough to compress the magnetospheres of Earth-like exoplanets there to the standof distance of 3 Earth radii. \\

Compression of the magnetosphere of Jupiter results in UV emission~\citep{Nichols2017} and radio emissions, which can therefore be used as a proxy for compressions~\citep{Desch1984,Lamy2012,Hess2014,Dunn2016}. These compressions can lead to atmospheric erosion and~\cite{Edberg2011} report that the atmospheric escape rate of Venus's atmosphere rose by 36\%. \cite{Dubinin2009} previously showed that Mars's ionosphere was impacted by a singular large CIR and resulted in an increase of a factor of 10 in the escape rate of the atmosphere.

\begin{figure*}
    \centering \includegraphics[width=1.8\columnwidth]{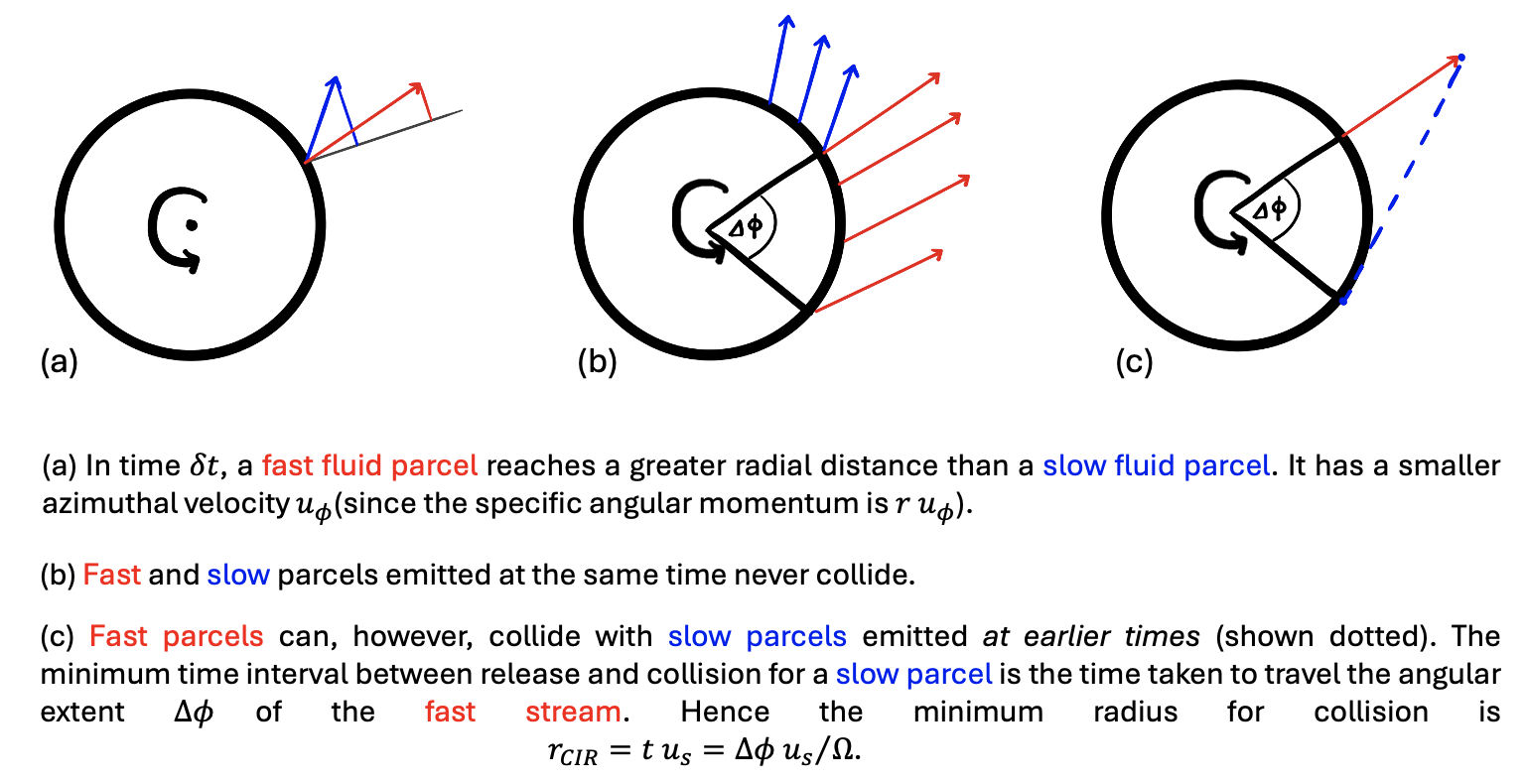}
    \caption{Schematic showing how the fast (red) and slow (blue) winds interact and the minimum radius, $r_{CIR}$, at which a CIR forms. The system is viewed from a point along the stellar rotation axis.}
\label{fig:cartoon2}
\end{figure*}

\subsection{Solar CIRs}
CIR shocks in our solar system typically form beyond 2 AU \citep{Gosling1976,Gazis1983} but they can have formed by 1 AU \citep[e.g.][]{Richardson1984,Berdichevsky2000}. A study of shocks from 1995-2004 \citep{Jian2006} reported that 17\% of CIRs had a forward leading shock and 6\% had a reverse shock at 1 AU. At closer distances to the Sun, the number of CIRs with shocks decreases \citep{Schwenn1990,RichardsonReview}. One study suggested that the rate of CIRs with shocks closer than 0.5 AU was about 1 every 200 days, whilst between 0.5-1 AU it was half this \citep{Richter1985}. At larger orbits within our solar system, the frequency of CIRs with associated shocks increases, with 90\% having forward shocks and 75\% with reverse shocks by 3-5 AU (~\citet{RichardsonReview} and references within). By about 10-12 AU, the shocks decline, although evidence of the CIRs still exists to 15-20 AU in the form of co-rotating pressure enhancements. As the CIR moves outward within the Solar System, it begins to expand and erode, although the regions of high magnetic field and plasma density remain (~\cite{RichardsonReview} and references within).
\\
Solar CIRs are associated with high energy particles in the form of ions and protons that are accelerated, primarily within the shocks~\citep{Cohen2020,RichardsonReview}. These particles typically have energies of around 10-20 MeV and although generally this is ascribed to acceleration within the shocks, there is also evidence of some acceleration within the interaction region itself. The shocks are not identical, and particles from the reverse shock can produce a harder spectrum alongside a greater number of alpha particles \citep{BarnesSimpson1976}. The energies of the particles accelerated at the CIR shocks are far smaller than the energies associated with CMEs, making them less of a threat to Earth's atmosphere~\citep{Richardson2004}. However, on other stars these energies may be considerably higher due to the stronger winds of these stars. 


\subsection{CIRs in other low mass star systems}

Whilst CIRs are well studied on the Sun, very little is known about these features in other star systems, or even on the young Sun itself. This work aims to use fundamental physics in order to extract trends in behaviours, such that the differences in CIRs across various low mass stars may be understood. For example, many young low mass stars are rapid rotators which should result in tighter Parker spirals and thus tighter CIRs. The CIRs themselves should also form closer to the star due to the faster rotation periods. Planets orbiting such stars might be buffeted by CIRs more frequently than they would at the same orbit within our solar system, where a CIR may form less often. This suggests that on some stars, CIRs may be as important as the more commonly-studied CMEs. MHD simulations suggest that on low mass M dwarfs, CME activity may be much less than on the present-day Sun, either because the weak surface differential rotation injects twist into flux tubes only very slowly \citep{Gibb2014,Gibb2016} or because CMEs once released, are confined by the strong magnetic field \citep{Alvarado2022}. Observations of stellar CMEs have also proved elusive \citep{Namekata2022,Inoue2023}. The winds of these stars may nonetheless still produce CIRs. On these stars the presence of CIRs may therefore be a significant, but thus far ignored, driver in the space weather in these systems. As mentioned above, CIRs on rapidly rotating stars might generate higher energy particles than is seen from the Sun. If these energies are high enough, this might lead to stripping of exoplanetary atmospheres \citep{2020IJAsB..19..136A}. \\
Hence, whilst we should expect that just like on the Sun, CIRs should be common occurrences across all low mass stars, their exact behaviour and importance may vary.


\section{Where do CIRs form around other stars?}
\label{sec:whereCIRsform}

As the wind flows outwards from the star, it carries the magnetic field with it. For a field that is radial at the stellar surface and in steady state, the magnetic field is driven into a Parker spiral \citep{Parker1958}. Both the fast and slow wind generate Parker spirals, but of different tightness due to their different velocities. Therefore, at some location around the star, these spirals will meet and the winds will interact, forming the inner point of the CIR. (see Figure~\ref{fig:cartoon1})\\

\begin{figure}
    \centering \includegraphics[width=1\columnwidth]{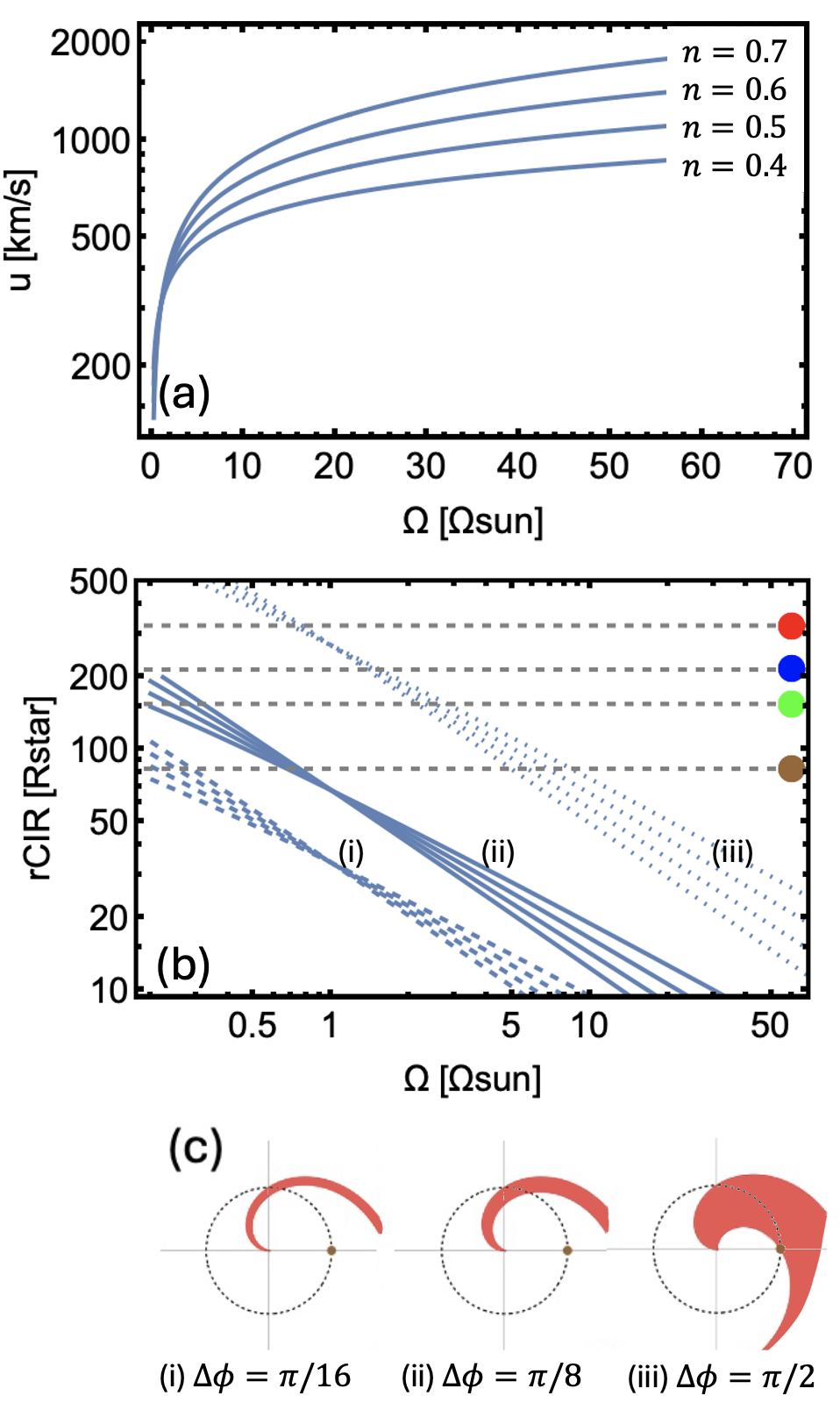}
    \caption{(a) A plot showing the wind velocity with rotation rate for the various power laws used (n = 0.4, 0.5, 0.6 and 0.7). (b) The minimum radius at which CIRs form, assuming a star of solar radius, with stellar rotation rate $\Omega$. The solid lines (ii) show curves for the angular extent of the fast wind stream $\Delta \phi = \pi/8$ and the dashed lines (i) for $\Delta \phi = \pi/16$. The dotted lines (iii) show the limiting case of $\Delta \phi = \pi/2$. In each case, curves are shown for the following power law relations in Equation \ref{eqn:temppowerlaw}; n = 0.4, 0.5, 0.6 and 0.7. The orbits of Mercury and Venus are shown by the brown and green points, respectively. Earth and Mars are shown by the blue and red points. (c) The effect of  $\Delta \phi$ on the fast wind stream.}
\label{fig:CIRplot1}
\end{figure}

Figure~\ref{fig:cartoon2}(a) shows vector descriptions of slow (blue) and fast (red) wind parcels, released from the same location on the stellar surface. The slow wind travels a shorter distance ($r$) in some time ($\Delta t$) than the fast wind, so the slow wind parcel can never ``catch up'' with the fast wind parcel. However, it has a larger $u_{\phi}$ than the fast particle (since the specific angular momentum is $r u_{\phi}$). Figure~\ref{fig:cartoon2}(b) shows fast parcels (red) released from within the fast wind segment (of angular width $\Delta\phi$), and three slow parcels being simultaneously emitted from further ahead on the surface. These fast parcels can never collide with the slow parcels as they have been emitted at the same time. However, the fast parcels can collide with slow parcels that were released at an earlier time (Figure~\ref{fig:cartoon2}(c)). In Figure~\ref{fig:cartoon2}(c), the fast parcel collides with a slow wind parcel that was released at an earlier time (blue dashed line). This slow parcel has travelled a distance $r = u_{slow} t$, where $t$ is the travel time of the slow parcel given by $t=\Delta\phi/\Omega$ (i.e. the angle the star has moved through divided by the angular rotation rate of the star). The radial location of the innermost point of the CIR is then:
\begin{equation}
    r_{CIR} = \Delta \phi\frac{u_{slow}}{\Omega}
    \label{eqn:rcir}
\end{equation}
where $\Delta\phi$ represents the angular separation of the fast and slow wind streams (depicted in Figure~\ref{fig:CIRplot1}(c)), $u_{slow}$ is the velocity of the slow wind stream and $\Omega$ is the stellar rotation rate. \\

The speed of the slow wind can be found using a thermal wind approximation~\citep{Parker1958,Blackman}, and thus it is dependent on temperature. At distances large compared to the location $r_{cs}=G M_{\star}/2c_s$ where the wind becomes supersonic ($u>>c_s$), we may write:
\begin{equation}
    u_{slow} \approx 2 c_s \sqrt{ \ln{r/r_{cs}}}
\end{equation}
where $c_s = \sqrt{K_b T/m}$ is the sonic speed. The full isothermal wind expression is given by
\begin{equation}
    \biggr(\frac{u}{c_s}\biggr)^2-ln\biggr(\frac{u}{c_s}\biggr)^2=4\biggr(\frac{r}{r_{cs}}\biggr)+4\biggr(\frac{r_{cs}}{r}\biggr)-3,
\end{equation}
given here for completeness \citep{Parker1958}.\\

Finally, the temperature can be related to the stellar rotation rate using a power law~\citep{Ivanova2003,Holzwarth2007,See2014,Ahuir2020}.
\begin{equation}
T = T_{\odot} (\Omega/\Omega_{\odot})^n.
\label{eqn:temppowerlaw}
\end{equation}

The slow wind speed is dependent on the temperature, in this approximation through the sonic radius and sound speed. Therefore, the wind speed is dependent on the stellar rotation rate (shown in Figure~\ref{fig:CIRplot1} (a) as the variation in wind velocity at 1 AU with stellar rotation rate). The sound speed is highly sensitive to rotation rate for slow rotators, however it is less sensitive at fast rotation rates. Figure~\ref{fig:CIRplot1} (a) shows the dependency for various values of $n$ in the power law (Equation~\ref{eqn:temppowerlaw}) and shows the increase in velocity for increasing $n$.\\

The minimum radius at which CIRs form is shown in Figure~\ref{fig:CIRplot1} (b). It can be seen that for the most rapidly rotating stars the minimum radius of CIRs does not change significantly with rotation rate. However, for the slowly rotating stars there is a steep decrease in CIR radius with increasing rotation rate. It is also clear that changing the power in the temperature-rotation rate relation does not significantly alter the results. The solid lines show results for the angular extent of the fast stream $\Delta \phi = \pi/8$ and dashed lines for $\Delta \phi = \pi/16$. The dotted lines show $\Delta \phi = \pi/2$, which is the maximum angle between the fast and slow wind for a dipole, and thus is the limiting case. In each of these cases, the four lines show results for different values of n (0.4, 0.5, 0.6, 0.7), i.e. different power-law relationships of coronal temperature with rotation rate. In this simple model with $\Delta \phi = \pi/8$, the closest CIR to the Sun could form at about 67.6 $R_{\odot}$ or 0.314 AU. For a simple model, this is close to what has observed of 0.3 AU \citep{Schwenn1990}. However, for a rapidly rotating star with a rotation period of 0.5 days (54$\Omega_{\odot}$), $r_{CIR}$ = 3.4 - 6.9 $R_{\odot}$, depending on the temperature power law used. This corresponds to 0.016 - 0.032 AU, a factor of 10 smaller than for the present-day Sun. For Mercury and Venus orbits, CIRs likely form inside of the planetary orbit for most stellar rotation rates (and therefore most stellar ages). However, there is some critical stellar rotation rate at which CIRs can no longer form within the planetary orbit. This critical rotation rate depends on the angular extent of the fast wind wedge, $\Delta \phi$, which we have treated as a constant here although it may vary with stellar rotation rate. However, for planets in Earth or Mars orbits, it may be the case that CIRs form within their orbits at all rotation rates, and therefore ages. With $\Delta \phi = \pi/2$, the lines for minimum CIR radius cross these planetary orbits, however this is not the case when $\Delta \phi = \pi/8$ is used. Thus, it may be the case that as a star ages, some planets experience a halting in the CIRs whilst for others they are a continuous aspect of stellar activity.\\

\begin{figure}
    \centering \includegraphics[width=0.9\columnwidth]{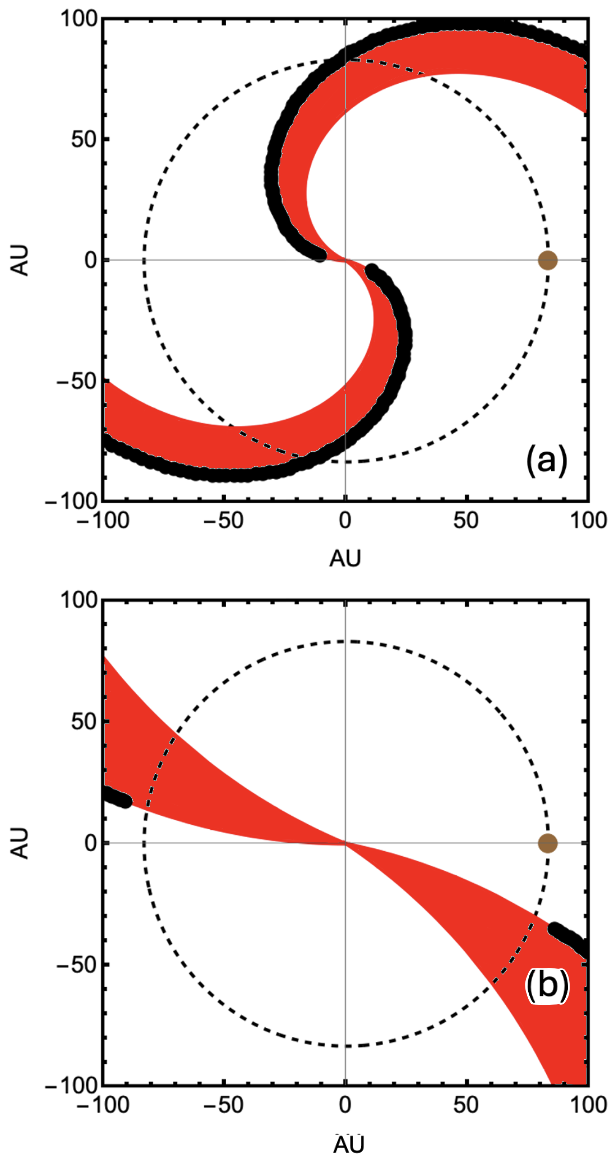}
    \caption{Figure showing the CIR streams for (a) a fast and (b) a slower rotating star. The red curves show the fast wind region embedded within the slower wind, whilst the locations of the CIRs are shown by the thick black lines. A Mercury orbit is shown by the black dashed line and brown point. The parameters used here are a $1M_{\odot}$, $1R_{\odot}$ star and the fast wind at 10 times the temperature of the slow wind, calculated from $T \propto \Omega^{0.6}$. (a) $P_{\star}=0.5$ days (b) $P_{\star}=15$ days.}
\label{fig:CIRspiral2}
\end{figure}

Although Equation~\ref{eqn:rcir} gives the minimum radius for a CIR, the overall shape simply follows the Parker spiral for the front edge of the fast wind stream. This is shown as the black curves within the plots in Figure~\ref{fig:CIRspiral2}. The red curves representing other fast stellar wind streamlines can be calculated from the Parker spiral:
\begin{equation}
   \phi = \phi_0 - \frac{\Omega}{u_{ft}} (r - 1) R_{\star},
\end{equation}
where $\phi$ represents the longitudinal coordinate of the streamline and $r$ represents the radial coordinate. $\phi_0$ represents the longitude where the streamline joins to the stellar surface, $\Omega$ is the stellar rotation rate and $u_{ft}$ is the velocity of the fast wind. Note here that $r$ is in units of stellar radii ($R_{\star}$). The black curve showing the CIR is thus plotted along the leading fast streamline, from the minimum radius found in Equation~\ref{eqn:rcir}.\\

Assuming a dipolar magnetic field structure embedded within the equatorial plane of the star, there will be two possible locations for CIR streams. The CIRs are shown by the thick black lines and the orbit of Mercury is shown by the black dashed circle and brown point. The CIRs begin at the radius given by Equation~\ref{eqn:rcir} and continue outwards along the leading fast stream line. The angular extent of the fast wind stream, $\Delta \phi$, is shown by the red shaded region. Figure~\ref{fig:CIRspiral2} shows solar-like stars at two rotation rates, (a) $P_{\star}=0.5$ days and (b) 15 days. The importance of rotation rate on the spiral tightness and minimum CIR radius is apparent here. Notably, for the rapid rotator the CIR forms within the planetary orbit, whilst for the slower rotator the CIR forms beyond the planetary orbit. The more rapid rotator also shows the tighter Parker spiral than is present for the slower rotator. In our solar system, high energy particles rain back towards Earth from CIR shocks further out. Planets orbiting a star with a tight spiral will be showered with particles for a larger proportion of their orbit than planets around slow rotators, where the spiral is far shallower. The flux of these particles will depend on the distance from the planet to the shock and given that the wind speed will be higher on faster rotators than slower ones, this means CIRs will steepen into shocks at smaller radii. Thus, planets around these stars could experience a higher bombardment of these high energy particles.

\begin{figure}
    \centering \includegraphics[width=0.9\columnwidth]{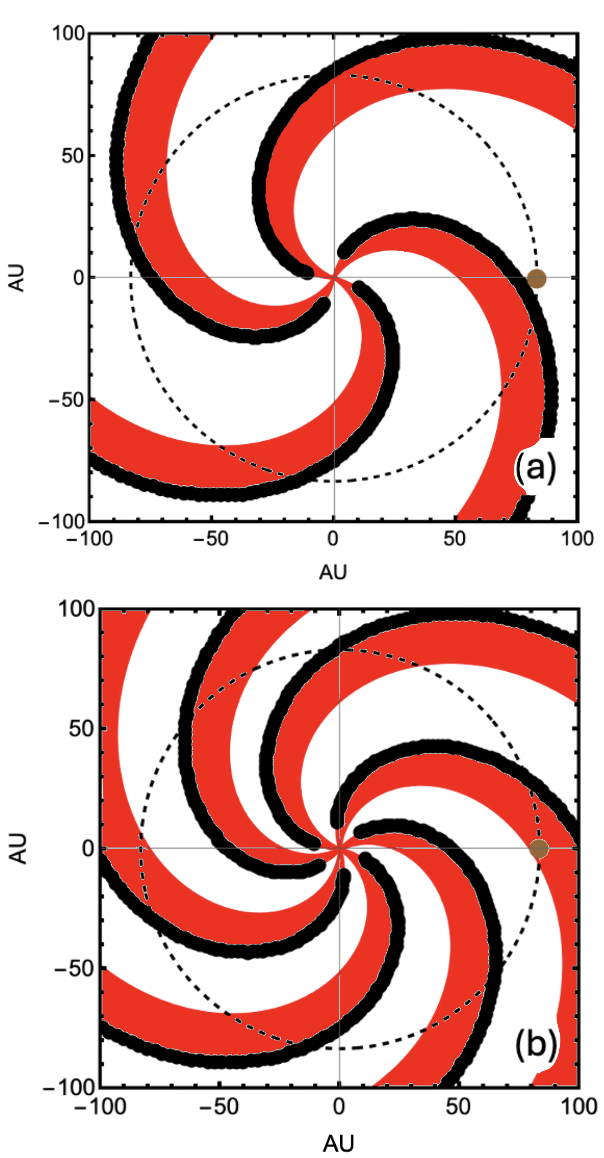}
    \caption{Same caption as Figure~\ref{fig:CIRspiral2}, but here (a) and (b) represent the quadrupole and octupolar fields, respectively.}
\label{fig:morecomplexfields}
\end{figure}

\section{Interactions with planets}

\subsection{Frequency of CIR-planet interactions}
The orbit of a planet around a star will govern how frequently it interacts with CIRs within the stellar system. However, the number of CIRs supported on a star, which depends on the field complexity, will also influence the frequency of interactions. The planet orbits the star with some frequency $f_{planet}$, and each CIR will co-rotate with the star i.e. have a rotational frequency of $f_{\star}$. Overall, the frequency of CIRs within the system will be given by $f_{CIR}=N f_{\star}$, where N is the number of CIRs supported by the star. The frequency of planet-CIR interactions can then be calculated from the beat frequency:
\begin{equation}
    f_{beat} = f_{CIR} - f_{planet}
    \label{eqn:f_beat}
\end{equation}
and, of course, $P_{beat} = 1/f_{beat}$.\\

\begin{figure}
    \centering \includegraphics[width=0.9\columnwidth]{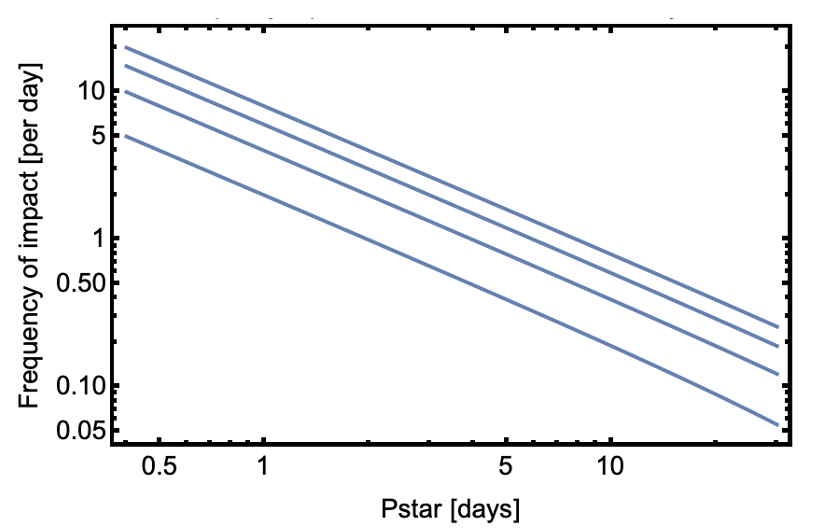}
    \caption{Frequency of CIR impacts for a Mercury orbit planet. The curves show impact frequency for systems with N = 2, 4, 6 and 8 CIR streams, from bottom to top of the plot respectively.}
\label{fig:planetinteractions}
\end{figure}

Whilst Figure~\ref{fig:CIRspiral2}(a) showed a star with 2 CIR streams, Figure~\ref{fig:morecomplexfields} shows a star with (a) 4 and (b) 6 CIR streams. This could represent, for example a (a) quadrupolar or (b) octupolar field structure, embedded within the equatorial plane of the star. The number of CIR streams present will determine the frequency of CIR impacts on orbiting planets, and this is shown in Figure~\ref{fig:planetinteractions} for a planet at Mercury's orbit. The bottom curve on the plot shows the frequency of impacts for a planet around a star with rotation period of half a day and 2 CIR streams. The curves above show 4, 6 and 8 streams, respectively. Planets around rapidly rotating stars experience far more frequent impacts than those around slower rotators, and the importance of the number of CIR streams on the impact frequency is larger for rapid rotators also. For the slower rotators, the frequency of impacts reaches an asymptote and is almost constant/independent of rotation rate. Here we have assumed that the CIRs are stable and present over the time periods here.\\

\begin{figure}
    \centering \includegraphics[width=1\columnwidth]{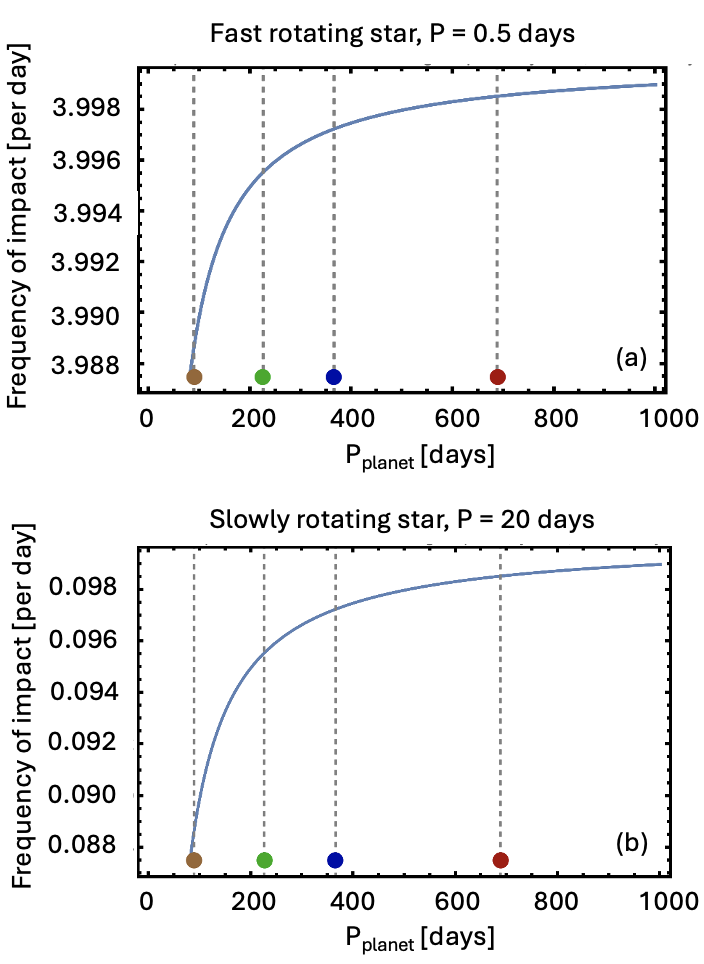}
    \caption{Frequency of planet-CIR interactions for a range of planetary orbits. Mercury, Venus, Earth and Mars are shown by the brown, green, blue and red points respectively, and the grey dashed lines. (a) Results for a star with a rotation period of 0.5 days and (b) a rotation period of 20 days.}
\label{fig:planetcir}
\end{figure}

The frequency of impacts can also be plotted against the planetary orbital period, which is shown in Figure~\ref{fig:planetcir}, assuming 2 CIR streams. Here, planets with Mercury (brown point), Venus (green point), Earth (blue point) and Mars (red point) orbits are shown for comparison. It is immediately clear that the orbital period of the planet has a far smaller influence on the frequency of CIR impacts than the stellar rotation period does. Figure~\ref{fig:planetcir}(a) shows the results assuming a rapid rotator with $P_{\star}=0.5$ days and (b) for a slow rotator with $P_{\star}=15$ days. In both cases, the frequency of impact does not vary significantly with orbital period. Whilst these plots show results for a 2 CIR stream system and adding more streams would increase the frequency of impact, the overall behaviour here would not change with more streams.



\subsection{Calculating the jump in ram pressure across the CIR}
\label{sec:CIRpressurejump}


While the previous sections consider the general trends of CIR location around planets, and the frequency of their interactions, this section considers the {\bf peak pulse in dynamic pressure} that a planet might experience. Before the CIR arrives, the ram pressure of the stellar wind at the planet's orbital radius is simply $p_{\rm {ram,before}} = \rho_1 u_s^2$ where $\rho_1$ is the density of the slow wind upstream of the CIR. Conditions behind the leading edge of the CIR can be determined from the Rankine-Hugoniot relations~\citep{Cravens1997}. The shocked gas has a density $\rho_2$ and a speed $U = 2(u_f-u_s)/(\gamma+1)$ (in the reference frame of the exoplanet). The planet therefore experiences a jump in dynamic pressure $\rho_2 U^2/\rho_1 u_s^2$. This can be rewritten as
\begin{equation}
\frac{p_{\rm {ram,after}}}{p_{\rm {ram,before}}} = 
    \frac{4(u_f/u_s - 1)^2}{(\gamma + 1)((\gamma-1) + 2/M_1^2)}
\end{equation}
where $\gamma$ is the ratio of specific heats and $M_1$ is the upstream Mach number given by $M_1 = (u_{fast} - u_{slow})/c_{slow}$, where $c_{slow}$ is the sound speed in the slow wind. In the limit where $M_1$ is large, this reduces to
\begin{equation}
\frac{p_{\rm {ram,after}}}{p_{\rm {ram,before}}} = 
    \frac{9}{4} (u_f/u_s-1)^2
\end{equation}
for $\gamma=5/3$. For typical solar wind values ($u_f \approx 750$kms$^{-1}$ and $u_s \approx 400$kms$^{-1}$) this gives a jump in dynamic pressure of about 1.7 - consistent with in-situ observations \citep{Edberg2011}.

\begin{figure}
    \centering \includegraphics[width=1\columnwidth]{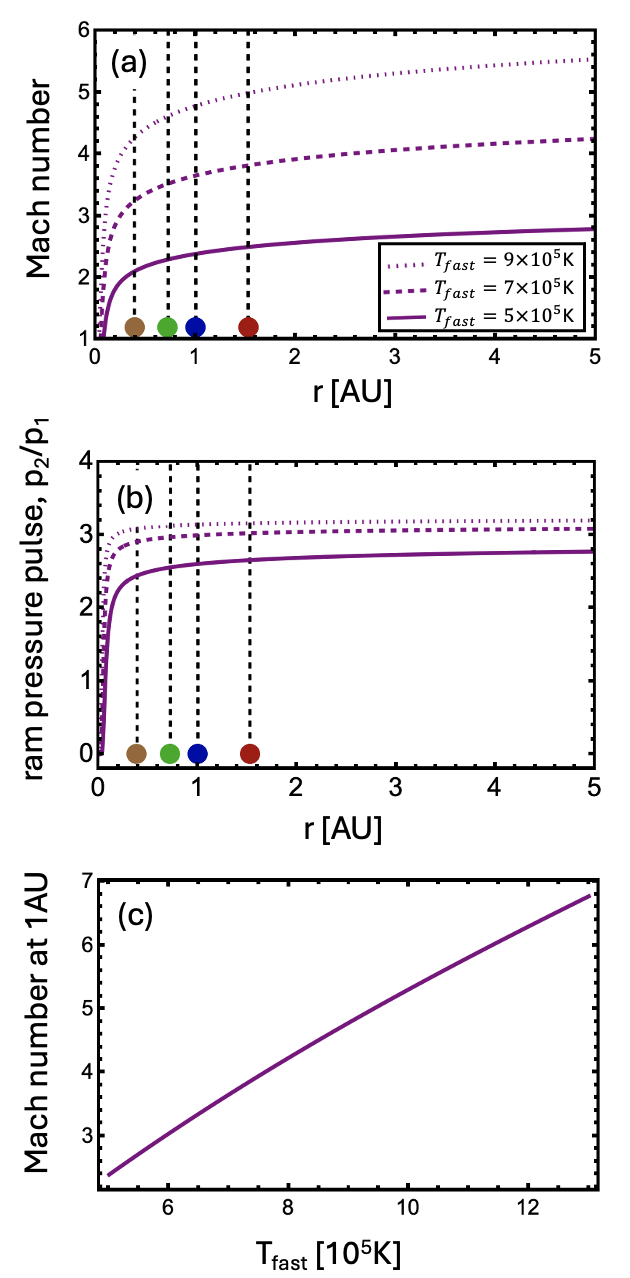}
    \caption{(a) Variation of CIR Mach number with radius from the star for parameters; $T_{slow} = 2\times 10^5$ K and $T_{fast} =$ 5, 7 and 9 $\times 10^5$ K. (b) Variation of the ram pressure jump within a CIR with radius from the star for parameters; $T_{slow} = 2\times 10^5$ K and $T_{fast} =$ 5, 7 and 9 $\times 10^5$ K. (c) The Mach number at 1AU for a slow wind of  $T_{slow} = 2\times 10^5$ K and varying fast wind temperatures.}
\label{fig:pressurestuff}
\end{figure}
For planets in large orbits, these velocities can be estimated as the terminal velocity of the winds as in Section~\ref{sec:whereCIRsform}. However, for the tight orbits considered in the remainder of this paper, where the wind may not have reached its terminal velocity, the velocities can be calculated from an isothermal wind. It is also noteworthy that as stellar mass ($M_{\star}$) decreases, the sonic radius moves in ($r_{cs} \downarrow$). For lower mass stars, the fast and slow wind speeds would both increase ($u_{f} \uparrow$ and $u_{s} \uparrow$). Thus, two stars with the same rotation rate but different masses will generate different CIR pressure pulses.\\

Figure~\ref{fig:pressurestuff} shows the Mach number and ram pressure jump that would be generated with a Parker wind for a slow wind of $T_{slow} = 2\times10^5$ K with a range of fast wind temperatures
($5\times10^5$ K (solid line), $7\times10^5$ K (dashed line) and $9\times10^5$ K (dotted line)). This pressure jump would be experienced as a pulse that sweeps past the planet as the CIR goes past. Planets in different orbits would experience different pressure pulses, and with the parameters $T_{slow} = 2\times10^5$ K and $T_{fast} = 5\times10^5$ K, a shock would be formed in any CIR forming above 0.2 AU, since the Mach number is already 1.8 at this radius. It is apparent that both the Mach number and pressure jump are sensitive to changes in the wind temperatures. The Mach number depends on the relative speeds of the winds, and thus in this case, the relative temperatures. Figure~\ref{fig:pressurestuff} (c) shows how the Mach number at 1 AU varies with temperature of the fast wind, holding the slow wind at $T_{slow} = 2\times10^5$ K. This shows an almost linear increase in Mach number with increasing fast wind temperature. Changes in temperature lead to a changes in the pressure jump, and since we have assumed that wind temperature is related to rotation rate (Equation~\ref{eqn:temppowerlaw}), the pressure jump can be related to this. 

\subsection{When might CIRs be important to planets?}

Both the magnitude of the ram pressure pulse and its frequency are important in considering the long-term consequences of the impact of CIRs. 
For planets with a magnetosphere, being buffeted regularly by CIRs at a large enough pressure could result in compression (followed by relaxation) of the magnetosphere. It follows that this could lead to heating and atmospheric loss \citep{2020IJAsB..19..136A}. Within our own solar system, in situ measurements allow us to quantify such losses, for both Mars \citep{2010GeoRL..37.3107E} and Venus \citep{Edberg2011}. While individual impacts by both CIRs and ICMEs (interplanetary coronal mass ejections) can result in an increase in ion loss of a factor of 10, a more typical figure is a factor of 2. The disturbance to the planetary outflow typically lasts for a period for around 4 days.

Placing this in the context of other stellar systems, we consider regular pressure pulses impacting an exoplanet and increasing its mass loss. The period between such pulses is $P_{beat}$, given by \ref{eqn:f_beat}. We denote the typical timescale of enhanced mass loss by $\Delta t$. The number of pulses per orbit is 
\begin{equation}
    n = \frac{P_{orbit}}{P_{beat}}.
\end{equation}
The fraction of each orbit during which the exoplanet suffers an enhanced mass loss is simply  $F=n \Delta t/P_{orbit}$, or
\begin{equation}
F = \frac{\Delta t}{P_{beat}}.
\end{equation}
Using \ref{eqn:f_beat} we can write this as
\begin{equation}
F = \frac{\Delta t}{P_{*}} 
       \left(N-\frac{P_{*}}{P_{orbit}} \right).
\end{equation}
For large orbital radii, where $P_{orbit}>> P_{*}$, this reduces to
\begin{equation}
F = \frac{N \Delta t}{P_{*}}.
\end{equation}
If the response time of the exoplanetary atmosphere ($\Delta t$) is greater than the time between pulses ($P_{beat}$) then the exoplanet is continually disturbed. In this phase, the exoplanet's atmosphere has no recovery time between pulses. This occurs at a critical stellar rotation period $P_{*,crit}$ given by
\begin{equation}
    \frac{P_{*,crit}}{P_{orbit}} \leq \frac{N \Delta t}{P_{orbit} + \Delta t}.
\end{equation}
In the limit of large orbital radii, where $P_{orbit}>> \Delta t$, this reduces to
\begin{equation}
    P_{*,crit} \leq N \Delta t.
\end{equation}
Placing this in context, if we use a typical solar system value of $\Delta t = 4$ days, then an exoplanet will be continually impacted by CIRs if there are 2 CIRs per orbit and the rotation period of the host star is less than 8 days.

\begin{figure}
    \centering \includegraphics[width=1\columnwidth]{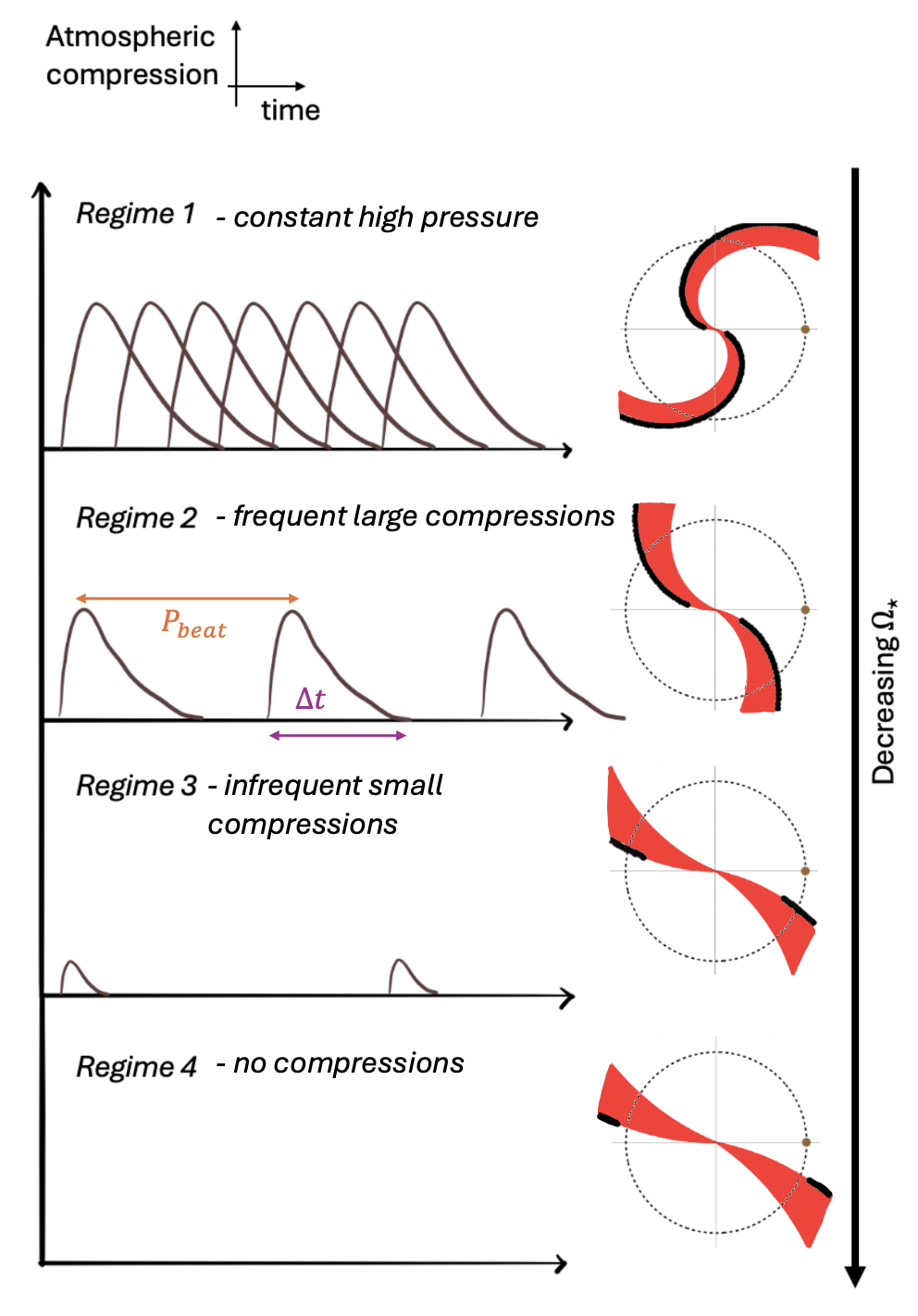}
    \caption{Conceptual depiction of various regimes that exoplanets may experience regarding CIR-planet interactions. The top panel shows the planetary atmospheric response during a CIR interaction for an ultrarotating star; regime 2 shows a fast rotator; regime 3 shows a slow rotator and regime 4 shows the situation where the CIRs always form beyond the planetary orbit.}
\label{fig:thwackingcartoon}
\end{figure}

Determining the outflow rate of charged ions that might result from such pulses requires a knowledge of the exoplanetary composition and the ambient loss rate. While these are not yet known for exoplanets, we can estimate the {\it relative} increase that such pulses might cause, using the approach in \citet{Edberg2011}. We write the ambient outflow rate as $\ell$ [m$^{-2}$$s^{-1}$]. This is enhanced by a factor $R$ during the passage of a CIR. During each orbit, the total outflow during a CIR is then $\ell R F$, and the total outflow when there is no CIR impact is $\ell(1-F)$. In each orbit, the outflow during CIRs as a fraction of the total is simply $RF/(1-F+RF)$ or
\begin{equation}
    \rm{Relative \ Outflow} =  \frac{R \Delta t}
        { 
         \frac{P_*}{(N-P_{*}/P_{orbit})} + (R-1)\Delta t 
         },
\end{equation}
which for large orbits reduces to
\begin{equation}
    \rm{Relative \ Outflow} =  \frac{R \Delta t}
        { P_*/N + (R-1)\Delta t }.
\end{equation}
For typical solar system values of $R=2$ and $\Delta t =4$days \citep{Edberg2011} we have
\begin{equation}
    \rm{Relative \ Outflow} =  \frac{8N}
        { P_*[\rm days] + 4N }.
\end{equation}
The relative outflow during CIRs therefore decreases as the stellar rotation period increases during the star's main sequence lifetime. This outflow falls to $50\%$ of the total at a rotation period given by $P_*[\rm days] = N\Delta t(R+1)$. We therefore expect that as a star ages and spins down, there is a systematic change in CIR-enhanced outflows from any orbiting exoplanets. We characterise these as follows (see  Figure~\ref{fig:thwackingcartoon}). 
\\

\textit{Regime 1 - constant high compression ($P_*\leq N\Delta t$):} For an ultrafast rotating star, an orbiting planet will be hit frequently (see Figure~\ref{fig:planetinteractions}), and this will be true for planets in all orbits since the stellar rotation period dominates over the planetary orbital period (see Figure~\ref{fig:planetcir}). i.e. both close in and far out planets will be regularly hit by CIRs. 
The planet will be hit so frequently that the planetary outflow may not have time to re-establish its previous equilibrium because the timescale of CIR impacts is less than the relaxation timescale for the outflow. The planet will experience an elevated outflow rate that is determined by the cumulative CIR impacts.\\
\textit{Regime 2 - frequent large compressions ($N\Delta t < P_* \leq N\Delta t(R+1)$)}: Around fast rotators, there may be time between impacts for the planetary outflow to recover, rather than being in a state of constant disturbance. In this regime, the impacts are still strong, but now the planet experiences regular changes in pressure, jumping from the ambient wind pressure to the CIR pressure as the CIRs sweep past. Thus, the planetary magnetosphere is periodically compressed and then relaxes, leading to planetary heating and potentially enhanced atmospheric loss. Whilst the CIRs are less numerous and powerful than in regime 1, they may be more consequential for the planet.\\
\textit{Regime 3 - infrequent small compressions ( $P_* > N\Delta t(R+1)$)}: For slowly rotating stars, the CIR impacts are small and less frequent. Here the CIRs are likely less consequential for the planet due to these combined factors.\\
\textit{Regime 4 - no compressions}: For some planets, there comes a stellar rotation rate at which CIRs form beyond the planetary orbit and thus the planet is no longer hit by CIRs.\\

Given that stars spin down with time on the main sequence, this means that planets will move from one regime to another as they age. However, some planets may not experience all of these regimes. Planets in close-in orbits such as Mercury and Venus could experience a different subset of these regimes than further out planets such as Earth or Mars. For a planet at a tight orbit such as Mercury, the critical stellar rotation period at which CIRs form beyond the planetary orbit will occur when the star is still quite young and thus the planet may skip regime 3. However, for an Earth orbit, the minimum CIR radius may always lie within the planetary orbit and thus such a planet would never experience regime 4.

\section{Conclusions}

The work here suggests that CIRs on rapidly rotating stars could be an important aspect of stellar activity. With a simple model, it can be shown that CIRs form at closer distances to these stars than is seen on the Sun. The closest CIR to the present-day Sun would form at about 0.314 AU (67.6 $R_{\odot}$), consistent with observations, and for a star with a rotation period of 0.5 days, $r_{CIR} = (0.016 - 0.032) $AU, i.e. (3.4 - 6.9 $R_{\odot}$). We list below the main conclusions:

\begin{itemize}
    
\item \underline{The minimum radius of CIRs} is dominated by the rotation rate and is not strongly affected by the coronal temperature relation, despite the dependence of $u_{slow}$ on temperature. The same behaviour was found with various power law scaling relations of coronal temperature to stellar rotation rate.\\

\item The \underline{frequency of CIR-planet interactions} is strongly tied to the stellar rotation period for the fastest rotators.  The frequency of impacts then tends towards an asymptote as rotation rate decreases, and is almost constant for the slowest rotators. The number of CIRs present in a system, which will depend on field geometry, will also influence the frequency of impacts. For the fastest rotators, the frequency of CIR impacts is almost constant from Mercury to Mars orbits.\\

\item The \underline{CIR-induced dynamic pressure pulse} experienced by a planet depends mainly on the 
\textit{difference} in wind velocities, and therefore the difference in wind temperatures. Large differences between the fast and slow wind temperatures lead to larger pressure jumps. 
\\

\item We have identified \underline{4 regimes which a planet may experience}, relating to CIR importance. These depend primarily on the number $N$ of CIRs around the star and the stellar rotation period $P_*$ (and hence the stellar and  exoplanetary ages). For an exoplanetary outflow that has a recovery time of $\Delta t$ and an increase by a factor $R$ in outflow rate during an impact, we identify the following regimes. \textit{Regime 1 ($P_*\leq N\Delta t$):} For planets around the most rapid rotators CIR impacts may occur on a shorter timescale than the timescale for the planetary outflow to relax. The planet will experience the cumulative effect of overlapping pressure pulses and an elevated outflow as it orbits the star. However, in this case the planet will likely not experience much variation in the dynamic pressure. \textit{Regime 2 ($N\Delta t < P_* \leq N\Delta t(R+1)$):} For fast rotators, the CIRs are less frequent but still strong, in which case the planet regularly experiences atmospheric compression and relaxation as the large CIR pressure pulse regularly sweeps past. Here CIRs are likely the most consequential to the planet. \textit{Regime 3 ( $P_* > N\Delta t(R+1)$):} For slow rotators the CIR impacts are less frequent and pressure pulses are smaller. Here the CIR direct impact is less important to the planet. Finally, in \textit{Regime 4} the planet no longer experiences direct collisions with CIRs as they always form beyond the planetary orbit. This does not mean that they are completely inconsequential, since further out CIRs in our Solar System are known to produce high energy particles that rain back towards the Earth and the other inner planets. However, here we have considered only the consequences of direct CIR-planet impacts. Planets may experience some but not all of these regimes, depending on how closely they orbit their parent star, and parent stars will evolve through various rotation rates as they age.\\

\end{itemize}

Overall, we have shown that CIRs on other low-mass stars should be frequent occurrences. The exact nature of these features in terms of frequency and strength is dependent on stellar rotation rate, whilst consequences for any orbiting planets will depend strongly on both the stellar rotation and the planetary orbit.

\section*{Acknowledgements}

The authors acknowledge support from STFC consolidated grant number ST/R000824/1.\\

The authors also thank the referee for the helpful comments that have improved the clarity of the manuscript.

\section*{Data Availability}
The research data from this paper can be accessed at: \url{ https://doi.org/10.17630/d7362bdc-dce6-46b4-b640-9360577ca10e}.\\

For the purpose of open access, the authors have applied a Creative Commons Attribution (CC BY) licence to any Author Accepted Manuscript version arising.



\bibliographystyle{mnras}
\bibliography{REF} 








\bsp	
\label{lastpage}
\end{document}